\documentclass[letterpaper,12pt]{article}
\usepackage{tabularx} 
\usepackage{amsmath}  
\usepackage{graphicx} 
\usepackage[margin=1in,letterpaper]{geometry} 
\usepackage{cite} 
\usepackage[final]{hyperref} 
\hypersetup{
	colorlinks=true,       
	linkcolor=blue,        
	citecolor=blue,        
	filecolor=magenta,     
	urlcolor=blue         
}

\usepackage{mathrsfs}
\usepackage{float}

\usepackage{revsymb}

\usepackage{amssymb}

\usepackage{exscale}

\usepackage{relsize}

\usepackage{gensymb}

\begin{document}

\title{Contribution of Wiggler to Radiation Integral\\ in an Electron Storage Ring}
\author{Xiujie Deng,\\ Tsinghua University, Beijing, China\\~\\
Ji Li, Yujie Lu,\\ Zhangjiang Laboratory, Shanghai, China}

\date{\today}
\maketitle

\begin{abstract}
With the advancement of accelerator light sources, the application of wiggler becomes more and more important, for example to speed up damping or generate synchrotron radiation. The quantum excitation contribution of such a wiggler to the electron beam emittance in a storage ring should be carefully evaluated when small emittance is desired. What found in literature is an approximate formula. Here we present a more exact result, which is of value for future light source development. 
\end{abstract}
\thispagestyle{empty}

This paper aims to give an accurate formula to evaluate the quantum excitation of a wiggler to beam emittance in an electron storage ring. The motivation is based on the fact that wiggler may become an essential part in future storage ring-based light source, for example to fight against intra-beam scattering by significantly increasing the natural radiation damping rate, where ultrasmall transverse or longitudinal emittance and high beam current is desired. Along with the radiation damping is the quantum excitation of the wiggler, which may not be negligible in determining the equilibrium beam emittance, especially when the wiggler field is strong and length is long. An approximate radiation integral formula can be found in literature to describe the wiggler's quantum excitation contribution. With the desired emittance becomes smaller smaller, an accurate formula which gives the exact dependence of the radiation integral on various optical functions at the wiggler is needed and here we fill this gap. 

In the following discussion, we focus on the case of a planar uncoupled ring which is the typical setup for present synchrotron radiation sources.  Note that the presented method however applies to a general coupled lattice~\cite{OSCSSMB2024}. Similarly, although we use horizontal plane for an example analysis, the method also applies to the vertical and longitudinal plane. Further, the method can also be used to evaluated the quantum excitation contribution of other bending-related elements.

Now we present the details. Usually the central part of the wiggler has a sinusoidal field strength pattern along the longitudinal axis $s$. We set $s=0$ to be the location of the peak magnetic field closest to the wiggler center. Then the vertical magnetic field of a horizontal planar wiggler is
\begin{equation}
B_{y}=B_{0w}\cosh(k_{x}x)\cosh(k_{y}y)\cos(k_{w}s),
\end{equation}
with $B_{0w}$ the peak magnetic field and $k_{w}=\frac{2\pi}{\lambda_{w}}$ the wavenumber of wiggler, and $k_{x}^{2}+k_{y}^{2}=k_{w}^{2}$.
Particle state vector 
$
{\bf X}\equiv\left(
\begin{matrix}
x&
x'& 
y&
y'&
z&
\delta
\end{matrix}
\right)^{T}
$ is used in this paper, with its components meaning horizontal location, angle, vertical location, angle, longitudinal location, and relative energy deviation with respect to the reference particle, respectively. $^{T}$ means transpose. The linear transfer matrix of ${\bf X}$ from $s=0$ to $s\in[-\frac{L_{w}}{2},\frac{L_{w}}{2}]$ with $L_{w}$ the wiggler length is then~\cite{Zhao2023}
\begin{equation}
	{\bf W}(s|0)=\left(
	\begin{matrix}
	
		1&s&0&0&0&-\frac{K}{\gamma k_{w}}[1-\cos(k_{w}s)]\\
		0&1&0&0&0&-\frac{K}{\gamma}\sin(k_{w}s)\\
			0&0&\cos(k_{y}s)&\frac{\sin(k_{y}s)}{k_{y}}&0&0\\
		0&0&-k_{y}\sin(k_{y}s)&\cos(k_{y}s)&0&0\\
		W_{51}&W_{52}&0&0&1&W_{56}\\
		0&0&0&0&0&1
	\end{matrix}
	\right)
\end{equation}
where $W_{51}=-W_{26}$, $W_{52}=-sW_{26}+W_{16}$ and 
$
W_{56}
=\frac{2\lambda_{0}}{\lambda_{w}}s+\frac{K^{2}}{\gamma^{2}}\left[\frac{\sin(2k_{w}s)-4\sin(k_{w}s)}{4k_{w}}\right],
$
with $\lambda_{0}=\frac{1+K^{2}/2}{2\gamma^{2}}\lambda_{w}$ being the fundamental on-axis resonant wavelength, $\gamma$ is the Lorentz factor and for our interested relativistic cases is much larger than 1, $K=\frac{eB_{0w}}{m_{e}ck_{w}}$ is the dimensionless undulator parameter of the wiggler, with $e$ the elementary charge, $m_{e}$ the electron mass, $c$ the speed of light in free space.

In a planar uncoupled ring, the normalized eigenvector of the storage ring one-turn map corresponding to the horizontal eigenmode at $s=0$ can be expressed as
\begin{equation}\label{eq:eigen}
{\bf E}_{I}(0)=\frac{1}{\sqrt{2}}\left(
\begin{matrix}	
\sqrt{\beta_{x0}}\\
\frac{-\alpha_{x0}+i}{\sqrt{\beta_{x0}}}\\
0\\
0\\
\frac{-\left(\alpha_{x0}D_{x0}+\beta_{x0}D_{x0}'\right)+iD_{x0}}{\sqrt{\beta_{x0}}}\\
0
\end{matrix}
\right)e^{i\Phi_{I0}},
\end{equation}
where $\alpha_{x0},\beta_{x0}$ are the Courant-Snyder functions and $D_{x0},D_{x0}'$ are the dispersion and dispersion angle corresponding to the horizontal plane at $s=0$, $i$ is the imaginary unit. 
Then the chromatic function $\mathcal{H}_{x0}$ at $s=0$ is
\begin{equation}
	\mathcal{H}_{x0}\equiv2|E_{I5}(0)|^{2}=\frac{D_{x0}^{2}+\left(\alpha_{x0}D_{x0}+\beta_{x0}D_{x0}'\right)^2}{\beta_{x0}}.
\end{equation}
The evolution of $\mathcal{H}_{x}$ from $s=0$ to $s\in[-\frac{L_{w}}{2},\frac{L_{w}}{2}]$  is
\begin{equation}
	\begin{aligned}
		\mathcal{H}_{x}(s)&\equiv2|E_{I5}(s)|^{2}=2|\left({\bf W}(s|0){\bf E}_{I}(0)\right)_{5}|^{2}\\
		&=\frac{(D_{x0}+W_{16}-W_{26}s)^{2}+\left[\alpha_{x0}(D_{x0}+W_{16}-W_{26}s)+\beta_{x0}(D_{x0}'+W_{26})\right]^2}{\beta_{x0}},
	\end{aligned}
\end{equation}
where $W_{ij}$ means the $i$-th row and $j$-th column  matrix term of ${\bf W}(s|0)$.
Put in the explicit expression of the wiggler matrix terms, we have
\begin{equation}\label{eq:Hx}
\begin{aligned}
\mathcal{H}_{x}(s)&=\frac{1}{\rho _w^2 k_w^4 \beta _{x0}}\left\{[D_{x0}\rho_{w}k_{w}^{2}+\sin \left(k_{w}s\right)k_{w}s+\cos(k_{w}s)-1]^2\right.\\
&\left.+[\beta _{x0}k_w\left(\rho_{w}k_{w}D_{x0}'- \sin \left(k_{w}s\right)\right)+\alpha_{x0}\left(D_{x0}\rho_{w}k_{w}^{2}+\sin \left(k_{w}s\right)k_{w}s+\cos(k_{w}s)-1\right)]^2\right\}.\\
\end{aligned}
\end{equation}
We assume that the quantum excitation contribution from the entrance and exit region of the wiggler, where the field strength in reality deviates from the ideal sinusoidal pattern, is much smaller than that of the central sinusoidal field region.  Then the quantum excitation of a wiggler to the horizontal beam emittance can be evaluated by the integral~\cite{Sands}
\begin{equation}
	\begin{aligned}
		I_{5w}&=\int_{-\frac{L_{w}}{2}}^{\frac{L_{w}}{2}}\frac{\mathcal{H}_{x}(s)}{|\rho(s)|^{3}}ds=\frac{1}{\rho_{w}^{3}}\int_{-\frac{L_{w}}{2}}^{\frac{L_{w}}{2}}\mathcal{H}_{x}(s)|\cos(k_{w}s)|^{3}ds,
	\end{aligned}
\end{equation}
where $\rho_{w}=\frac{\gamma m_{e}\beta c}{eB_{0w}}$ corresponds to the bending radius at the location of peak magnetic field $B_{0w}$, with $\beta=\sqrt{1-\frac{1}{\gamma^{2}}}$.
Put Eq.~(\ref{eq:Hx}) in, we have
\begin{equation}\label{eq:I5wmost}
\begin{aligned}
I_{5w}
&=\frac{4}{15\pi}\frac{ L_{w}}{\rho_{w}^{5}k_{w}^{2}}\left[\beta_{x0}+\gamma_{x0}L_{w}^{2}\mathcal{R}+5\rho_{w}^{2}k_{w}^{2}\mathcal{H}_{x0}-\left(10+\frac{15\pi}{8}\right)\frac{\rho_{w}\left(D_{x0}+\alpha_{x0}D_{x0}+\beta_{x0}D_{x0}'\right)}{\beta_{x0}}\right],
\end{aligned}
\end{equation}
where $\gamma_{x0}=\frac{1+\alpha_{x0}^{2}}{\beta_{x0}}$ and
\begin{equation}\label{eq:correction0}
\mathcal{R}=\frac{15}{32\pi^{2} }\frac{1}{N_{w}^{3}}\int_{-N_{w}\pi}^{N_{w}\pi}[\sin \left(x\right)x+\cos(x)-1]^2|\cos(x)|^{3}dx,
\end{equation}
with $N_{w}=\frac{L_{w}}{\lambda_{w}}$ the number of wiggler period which is assumed to be an integer. Equation~(\ref{eq:I5wmost}) above is the exact formula for the radiation integral $I_{5}$ of a wiggler in an electron storage ring.

Given a specific $N_{u}$, $\mathcal{R}$ can be straightforwardly obtained by integration in Eq.~(\ref{eq:correction0}).  When $N_{w}\gg1$, we have  
$
\mathcal{R}\approx\frac{1}{12},
$
and
\begin{equation}\label{eq:I5waccurate}
\begin{aligned}
I_{5w}
&\approx\frac{4}{15\pi}\frac{ L_{w}}{\rho_{w}^{5}k_{w}^{2}}\left[\langle\beta_{x}\rangle_{w}+5\rho_{w}^{2}k_{w}^{2}\mathcal{H}_{x0}-\left(10+\frac{15\pi}{8}\right)\frac{\rho_{w}\left(D_{x0}+\alpha_{x0}D_{x0}+\beta_{x0}D_{x0}'\right)}{\beta_{x0}}\right],
\end{aligned}
\end{equation}
where $\langle\beta_{x}\rangle_{w}$ is the average $\beta_{x}$ along the wiggler
\begin{equation}
\langle\beta_{x}\rangle_{w}=\frac{1}{L_{w}}\int_{-\frac{L_{w}}{2}}^{\frac{L_{w}}{2}}\beta_{x}(s)ds=\beta_{x0}+\frac{\gamma_{x0}L_{w}^{2}}{12}.
\end{equation}
Denote $\chi_{x0}=\text{arg}\left(\frac{E_{I5}(0)}{E_{I1}(0)}\right)$, where $E_{I5}(0)$ and $E_{I1}(0)$ represent the fifth and first term of the eigenvector Eq.~(\ref{eq:eigen}) and arg() means the angle of a complex number, then
Eq.~(\ref{eq:I5waccurate}) can be written as
\begin{equation}\label{eq:I5wNew2}
\begin{aligned}
I_{5w}
&\approx\frac{4}{15\pi}\frac{ L_{w}}{\rho_{w}^{5}k_{w}^{2}}\left[\langle\beta_{x}\rangle_{w}+5\rho_{w}^{2}k_{w}^{2}\mathcal{H}_{x0}-\left(10+\frac{15\pi}{8}\right)\rho_{w}\sqrt{\frac{2\mathcal{H}_{x0}}{\beta_{x0}}}\sin\left(\chi_{x0}-\frac{\pi}{4}\right)\right].
\end{aligned}
\end{equation}
The first term in the above bracket corresponds to the approximate formula found in literature
\begin{equation}
I_{5w,\text{intrinsic}}\approx\frac{4}{15\pi}\frac{ L_{w}\langle\beta_{x}\rangle_{w}}{\rho_{w}^{5}k_{w}^{2}}.
\end{equation} 
It can be viewed as the intrinsic contribution of a wiggler to the radiation integral $I_{5}$, since there will be intrinsic dispersion and dispersion angle generated inside the wiggler even if $D_{x0}=0$ and $D_{x0}'=0$ as can be seen from the matrix term $W_{16}$ and $W_{26}$ of the wiggler. The second and third term in the bracket arise from a nonzero $\mathcal{H}_{x0}$. When $D_{x0}'$ or $\frac{D_{x0}}{\beta_{x0}}$ is of the order $\frac{K}{\gamma}=\frac{1}{\rho_{w}k_{w}}$, which can easily be the case in a real lattice, the contribution from this nonzero $\mathcal{H}_{x0}$ in the bracket could be comparable or even larger than the first term and cannot be neglected. The more accurate formula derived here should then be invoked to calculate the wiggler's quantum excitation of beam emittance. When the third term is much smaller than the second term, i.e., roughly when $\mathcal{H}_{x0}\beta_{x0}\gg\left(\frac{K}{\gamma}\lambda_{w}\right)^{2}$, Eq.~(\ref{eq:I5wNew2}) can be further approximated as
\begin{equation}\label{eq:I5wNew3}
\begin{aligned}
I_{5w}
&\approx\frac{4}{15\pi}\frac{ L_{w}\langle\beta_{x}\rangle_{w}}{\rho_{w}^{5}k_{w}^{2}}+\frac{4}{3\pi}\frac{ L_{w}\mathcal{H}_{x0}}{\rho_{w}^{3}},
\end{aligned}
\end{equation}
where the first term accounts for the intrinsic contribution, and the second term for the nonzero $\mathcal{H}_{x0}$.


From the above analysis, we can see that the minimum $I_{5w}$ is realized when
\begin{equation}
\alpha_{x0}=0,\ \beta_{x0}=\frac{L_{w}}{2\sqrt{3}},\ D_{x0}=0,\ D_{x0}'=0,\ \left(\mathcal{H}_{x0}=0\right),
\end{equation}
and the minimal value is
\begin{equation}
I_{5w,\text{min}}\approx\frac{4}{15\sqrt{3}\pi}\frac{L_{w}^{2}}{\rho_{w}^{5}k_{w}^{2}}.
\end{equation}

We believe the calculation method and analysis presented here can be useful in optimizing wiggler in future accelerator light source to realize an ultrasmall beam emittance.

\end{document}